# A Way Forward for Cosmic Shear: Monte-Carlo Control Loops

Alexandre Refregier[*] & Adam Amara[†]
*ETH Zurich, Department of Physics, Wolfgang-Pauli-Strasse 27, 8093, Zurich, Switzerland*

**Abstract**

Weak lensing by large scale structure or 'cosmic shear' is a potentially powerful cosmological probe to shed new light on Dark Matter, Dark Energy and Modified Gravity. It is based on the weak distortions induced by large-scale structures on the observed shapes of distant galaxies through gravitational lensing. While the potentials of this purely gravitational effect are great, results from this technique have been hampered because the measurement of this weak effect is difficult and limited by systematics effects. In particular, a demanding step is the measurement of the weak lensing shear from wide field CCD images of galaxies. We describe the origin of the problem and propose a way forward for cosmic shear. Our proposed approach is based on Monte-Carlo Control Loops and draws upon methods widely used in particle physics and engineering. We describe the control loop scheme and show how it provides a calibration method based on fast image simulations tuned to reproduce the statistical properties of a specific cosmic shear data set. Through a series of iterative loops and diagnostic tests, the Monte Carlo image simulations are made robust to perturbations on modeling input parameters and thus to systematic effects. We discuss how this approach can make the problem tractable and unleash to full potential of cosmic shear for cosmology.

## 1. Introduction

Cosmology has made remarkable progress in recent decades thanks to the advent of new telescopes and instruments. This has led both to the confirmation of the ΛCDM cosmological model and to some of the most profound questions in fundamental science today: what is the nature of Dark Energy and Dark Matter, two mysterious components which together make up about 96% of the energy density of the universe today? What are the initial conditions that seeded the formations of structure in the Universe? Does Einstein's theory of gravity, general relativity, need to be revised on cosmological scales?

In order to answer these fundamental questions, a number of new experiments are coming online or are being planned. They are based on a combination of different cosmological probes, each of which gives a handle on different aspects of the model. One of these probes is weak gravitational lensing or 'cosmic shear' [1][2][3]. It is based on the measurement of the weak distortions arising from the bending of light by large-scale structures of the Universe on the observed shapes of distant galaxies. It has a special place in cosmology as it is able to map the distribution of Dark Matter in the Universe in 3-dimensions without making assumptions about the relationship between mass and light, as is needed for the other probes. Being a purely gravitational effect, it also gives a special handle on potential modifications of gravity. In terms of statistical errors, it is also potentially the most powerful probe to measure the properties of Dark Energy and its evolution, as described by both Dark Energy Task Force (DETF) [4] and the ESO-ESA Working Group on Fundamental Cosmology (WGFC) [5].

While the promises of cosmic shear are great, results from this technique have been slow to come. In spite of swift progress in the detection of the effect and during the first measurements in the early 2000s, measurements have been difficult and their impact on cosmology constraints (see eg. [6] for a recent compilation) limited by systematic effects. In the following we describe the origin of the problem and a way forward to bring about the full potential of cosmic shear, drawing upon approaches widely used in Particle Physics and Engineering.

---

[*] alexandre.refregier@phys.ethz.ch

[†] adam.amara@phys.ethz.ch

## 2. The Challenge

While the physics of cosmic shear is well understood and clean, the challenge in this technique lies in the difficult nature of the measurement. In particular, a demanding step in the cosmic shear analysis is the measurement of the shapes of faint galaxies from wide field CCD images. To reach a precision of a few percent on the equation of state of dark energy, a standard figure of merit for cosmological surveys, the ellipticities (or axis ratios) of galaxies must be measured with a precision of 1 part in four thousand [7]. This is made difficult by various systematic effects induced by the low signal to noise of the galaxies, the need to deconvolve the Point Spread Function of the instrument and other instrumental effects (see [8] and reference therein). In particular recent studies [9] have shown that noise bias, which is a second order noise term in the shape measurement process, is a serious limitations for shape measurements of galaxies with signal-to-noise ratios of roughly 10, which is typically used for weak lensing analyses.

Until now, the focus of research to get around this problem has been to develop general shear measurement methods. Several community-wide challenges have been set up (STEP and GREATs [10][11][12]) and have led to an improvement in the methods for measuring galaxy shapes (see [13] for discussion). However, the precision achieved in fully realistic conditions is still not sufficiently robust for future, and possibly current, surveys. For instance, a recent study [14 and reference therein] has explored in detail biases that can arise in the shape measurement process and shown that such terms, and the interplay between them, can be significant for future experiments.

## 3. A Way Forward

To get around this limitation and address the difficult problem of weak lensing shape measurements, a new approach is thus needed. An emerging and promising approach is to use image simulations as part of the shear calibration process [15]. We propose to build on this development and rely centrally on Monte Carlo simulations, as is done in other areas such as in Particle Physics experiments. In this approach, the forward measurement process is modeled via simulations that are repeated to average over the space of possible experimental configurations. To validate the simulations, we build a calibration framework based on control loops inspired from Engineering, where tolerance analyses, system level architectures and detailed preplanned calibration programs are done routinely, with impressive results[‡]. In contrast with general shape measurement methods, a key feature to this approach is that the system only needs to be customized and validated for a specific cosmic shear data set and instrument. The following describes the general scheme of the proposed calibration system. Features of possible implementations of the method is provided in Appendix A, while details of a specific implementation of this method to the Dark Energy Survey experiment [19] will be describe in a future paper [20].

## 4. Monte Carlo Control Loops

Figure 1 summarises a system-level scheme for building the calibrating process to measure the weak lensing signal of galaxies with high precision. At the heart of the approach is a reliance on Control Loops and Monte-Carlo methods to build the simulation infrastructure and to determine the appropriate level of complexity needed for the particular data set considered.

---

[‡] For example, see [16][17] for an introduction to Control and Feedback theory in Engineering and [18] for a discussion of ways of testing credit risk in financial engineering by varying input assumptions of stochastic simulations.

The data set (see box $\alpha_3$ in the figure) may contain not only the primary data, such as the main lensing survey, but also additional calibration data, such as subsamples of the data with deeper imaging, imaging in additional bands or time-domain images, and external data sets, such as high-density star fields and deep imaging of additional extra-galactic fields with higher angular resolutions.

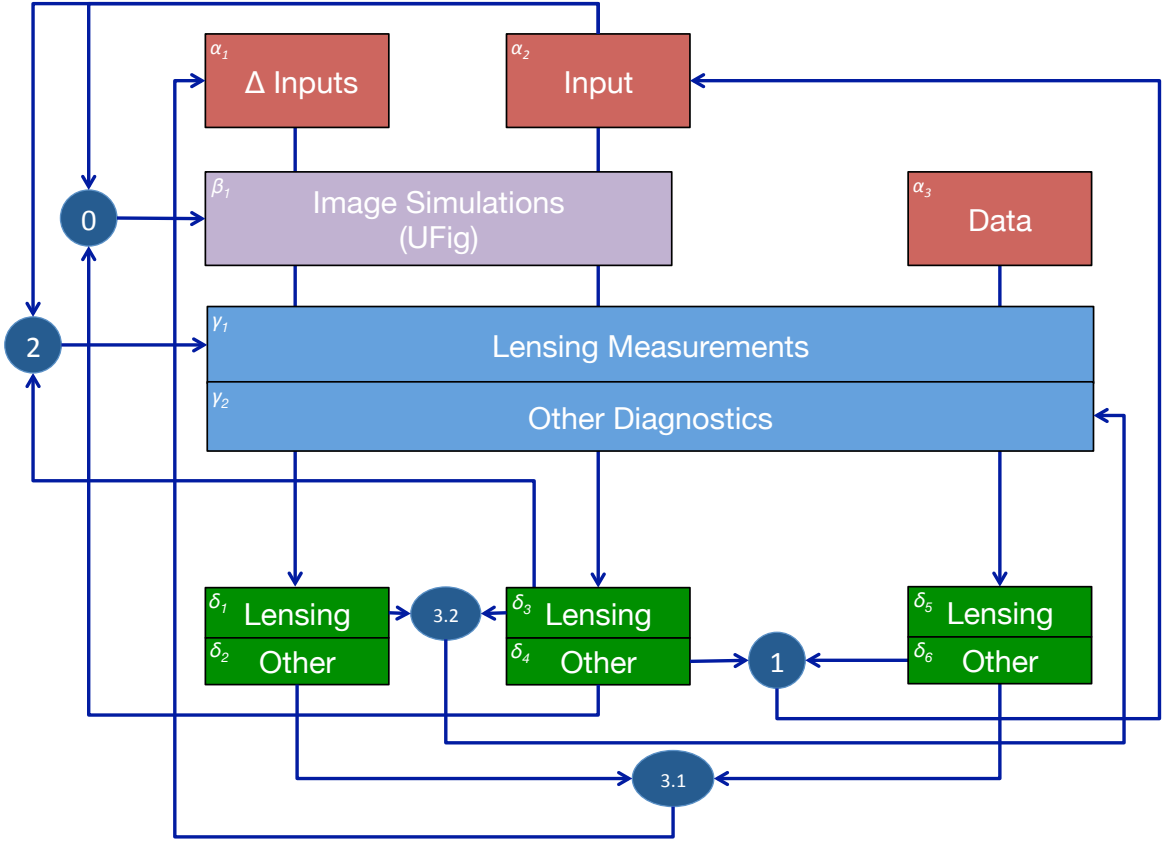

**Figure 1:** Schematic of the proposed Monte Carlo Control Loop system for the calibration of cosmic shear measurements customized for a specific data set. Red (labeled as α), purple (β), blue (γ) and green boxes (δ) represent inputs, simulations, measurements and outputs, respectively. Dark blue ellipses represent control loop tests, while arrows represent data and control flow.

This data will be processed through various processing and measurement analysis algorithms. These include lensing analysis algorithms ($\gamma_1$), which are used for cosmic shear, and other diagnostics measurements ($\gamma_2$), such as the magnitude-size distribution of galaxies, or PSF shape statistics as measured from stars. These produce output measurements ($\delta_5$ and $\delta_6$) that can then be used to derive the scientific results such as cosmological parameters.

The control loops shown in Figure 1 show a systematic scheme that can be used to (i) calibrate the lensing measurement stage, (ii) build up the necessary complexity in modeling and diagnostic tools; and (iii) identify the sources of uncertainty that have a significant impact on the shear measurement process. For this purpose, we place image simulations at the core of this infrastructure (see purple box $\beta_1$). As we show below, the scheme relies on a large number of iterations and so the speed of this simulation tool is paramount. In Figure 1 we identify the image simulation tool as UFig [21] (the Ultra-Fast Images simulator), which was developed with speed at the forefront so as to allow for a large number of control loops, however other image simulation tools may be used.

*Control Loop 0:* The purpose of this loop is to develop and test the image simulator. This is done by comparing the measured (non-lensing) output diagnostics ($\delta_4$) from the simulated images to the simulation inputs ($\alpha_2$). The loop is repeated by updating the simulations until these outputs are consistent. Note that this step assumes that the (non-lensing) diagnostics ($\delta_4$) are sufficiently reliable to provide a test of the simulator. This would be the case if they are based on widely used and well-tested codes such as Sextractor [22] applied to high signal-to-noise objects, but it may also require the inclusion of further diagnostics.

*Control Loop 1:* This loop is designed to tune the input parameters of the image simulator so that simulations are statistically consistent with the data. This is done by comparing the (non-lensing) diagnostic test outputs from the simulations and the data ($\delta_4$ and $\delta_6$). If they disagree, the input parameters of the simulation ($\alpha_2$) are modified and the loop is repeated. Since many of the steps in the analysis can be non-linear and unstable in the low signal to noise regime used in weak lensing, it is important that both the simulations and the real data are analysed in exactly the same way. A natural issue that arises at this stage is to identify what diagnostic tests are needed. Our view here is that initially the diagnostic tools should be minimal and simple since we will see below that we can implement a mechanism (control loop 3.2) to dynamically add complexity until we reach the precision that we need to fully exploit the data.

*Control Loop 2:* This step calibrates the lensing measurement method. This is done by varying the input lensing signal ($\alpha_2$) in the simulations and by comparing this to output lensing measurements ($\delta_3$). Calibration parameters in the lensing measurement algorithm ($\gamma_1$) are varied and the loop is repeated until these match. In practice this step, or something similar, is performed by most measurements methods that have been used [see 13 and reference therein]. The key difference here is that instead of being done in an ad-hoc way, this process is explicitly built-in as an integral part of the system architecture.

*Control Loop 3.1:* The control loops at level 3 introduce Monte-Carlo methods to test the robustness of the calibration scheme. The first step involves changing the input parameters about the fiducial values from Loop 2 above ($\alpha_1$) and checking that the output ($\delta_2$) of the new simulations are still consistent with the data ($\delta_6$). If outputs for the new simulations and the data are not in agreement, then a new set of Delta inputs is drawn until a new simulation configuration is found that passes the tests set by the diagnostics.

*Control Loop 3.2:* For all simulation inputs that pass the test in Loop 3.1, we need to ensure that the fiducial calibration method from loop 2 is valid and stable. Should we find cases that satisfy our diagnostic test (loop 3.1 and thus loop 1) but require a different shear calibration scheme (i.e. different outputs for the green boxes on the left and center), then we would know that our calibration method is not robust over all changes of the inputs consistent with the data. The way to resolve this problem would then be to make the diagnostic tests more stringent. This would thus reduce the space of plausible inputs such that the calibration scheme is stable over this space. Once additional diagnostic tests have been added, the whole system should be restarted from Loop 1 and the iterations should continue until the results from Loop 3.2 remain stable. If this process does not converge, this would mean that there is not sufficient information in the data to calibrate the measurement. In this case more data is needed, as for example a high resolution data set from space (eg. HST) if ground based data is being analysed. Once this new data set has been added the process needs to be started again from loop 1.

Once this iterative process has been completed, the measurement process will have been calibrated and tested for robustness to systematic errors. We can then proceed to measuring the lensing signal from the data and infer cosmological information. Note that this control loop process requires a large number of iterations over large simulated data sets and is thus facilitated by fast image generators such as UFig. Note also that, as for any other measurement process, potentials unknown systematics may affect the measurement. However, the proposed approach provides a framework for testing any aspect of the measurement process that is in doubt.

## 5. Conclusion

If cosmic shear is to live up to its promise and deliver enough to justify the large resources invested in this field, we advocate that a new approach is needed. We thus propose the Monte Carlo Control Loop scheme described above as a way to calibrate the shear measurement for specific data sets by integrating practices that are common in Particle Physics and Engineering. We believe that by viewing the entire measurement and calibration process as a global system that this problem can be made tractable. Cosmic Shear should then be able to lead the way in the coming era of cosmic discoveries and shed new light on some of the Universes deepest secrets in the coming decade.

**Acknowledgements**

We are grateful to Joël Bergé and Lukas Gamper for stimulating discussion at initial stages of this work and during the development of UFig. We also thank Sarah Bridle for useful discussions and Bhuvnesh Jain and Gary Bernstein for their comments.

**Appendix A: Implementation**

In this appendix, we describe features of possible implementations of the proposed scheme. A detailed description of the specific implementation of the method to the Dark Energy Survey [19] will be described in a future paper [20].

As explained in Section 4 above, the control loops consists of a large number of iterations of simulated images with various values of simulation parameters to which image processing analyses are applied.

The large number of iterations necessary for the control loop process is made possible by the Ultra Fast Image Generator UFig described in [21]. The number of iterations will thus be limited by the speed of the generation of a UFig image and by the data processing time to produce the diagnostics ($\gamma_1$ and $\gamma_2$ in Figure 1). As we showed in [21], the UFig generation of a Subaru of image of 0.25 square degrees (with 10k x 8k pixels with a limited magnitude of R≈26 takes about 30 sec using 4 cores on a current laptop such as a macBook Pro with a 2.7 GHz Intel processor). Assuming a comparable time for the data analysis of the image with Sextractor, this means that we can generate and analyse 800 Subaru images (200 deg$^2$) in 30 minutes with 100 cores. Assuming that an image model has 10 parameters, which is plenty for the cases we have looked at, and that we want to explore 10 values for each of these parameters about a fiducial model at the loop 3 level (the most time consuming loop in terms of iterations), we would need 100 iterations in this loop which can be done in 2 days with 100 cores. Since current surveys are about 200 deg$^2$ and cluster resources at the level of hundreds of cores are readily available, we see that it is feasible to perform this calibration on the time scale of a few days. For future larger surveys, the computation time will scale approximately like the survey area, thus requiring either faster or more numerous cores or longer computations.

At each iterations, to test whether the simulated images are consistent with the data we will follow the following steps, we will first run the same image analysis algorithm (eg. Sextractor) on both simulated and real images. We will then apply the same set of diagnostics to both which will typically be in the form of 1-dimensional (ex. Pixel intensity or magnitude distributions) or 2-dimensional histograms (eg. 2 component ellipiticity distributions, or size-magnitude distributions). We will then apply either a 1D or 2D Kolmogorov-Smirnov test [23] or a Chi-square test on these binned distributions. This will give the likelihood that the object catalogues of the data and of the simulations are drawn from the same distribution.